\documentclass[%
reprint,
amsmath,amssymb,
aps,
]{revtex4-1}

\usepackage{graphicx}% Include figure files
\usepackage{dcolumn}% Align table columns on decimal point
\usepackage{bm}% bold math
\usepackage{gensymb} 
\usepackage[hidelinks]{hyperref}% add hypertext capabilities
\usepackage{xcolor}
\begin{document}
	
	\preprint{APS/123-QED}
	
	\title{Theoretical insights of codoping to modulate electronic structure of TiO$_2$ and SrTiO$_3$ for enhanced photocatalytic efficiency}% Force line breaks with \\
	%\phone{+91-2659 1359}
	%\fax{+91-2658 2037}
	\author{Manish Kumar}
	\email{manish.kumar@physics.iitd.ac.in}
	\author{Pooja Basera}
	\author{Shikha Saini}
	\author{Saswata Bhattacharya}
	\email{saswata@physics.iitd.ac.in}
	\affiliation{Department of Physics, Indian Institute of Technology Delhi, New Delhi 110016, India\\}
	%\affiliation{Department of Physics, Indian Institute of Technology Delhi\\  Hauz Khas, New Delhi 110016,India.}%
	
	%\collaboration{MUSO Collaboration}%\noaffiliation
	
	%\author{Charlie Author}
	% \homepage{http://www.Second.institution.edu/~Charlie.Author}
	%\affiliation{ Second institution and/or address\\This line break forced% with \\}%
	%\affiliation{Third institution, the second for Charlie Author}%
	%\author{Delta Author}
	%\affiliation{%
	% Authors' institution and/or address\\
	%This line break forced with \textbackslash\textbackslash}%
	
	%\collaboration{CLEO Collaboration}%\noaffiliation
	
	\date{\today}% It is always \today, today,
	%  but any date may be explicitly specified
\begin{abstract}
	\noindent
TiO$_2$ and SrTiO$_3$ are well known materials in the field of photocatalysis due to their  exceptional electronic structure, high chemical stability, non-toxicity and low cost. However, owing to the wide band gap, these can be utilized only in the UV region. Thus, it's necessary to expand their optical response in visible region by reducing their band gap through doping with metals, nonmetals or the combination of different elements, while retaining intact the photocatalytic efficiency. We report here, the codoping of a metal and a nonmetal in anatase TiO$_2$ and SrTiO$_3$ for efficient photocatalytic water splitting using hybrid density functional theory and \textit{ab initio} atomistic thermodynamics. The latter ensures to capture the environmental effect to understand thermodynamic stability of the charged defects at a realistic condition. We have observed that the charged defects are stable in addition to neutral defects in anatase TiO$_2$ and the codopants act as donor as well as acceptor depending on the nature of doping (p-type or n-type). However, the most stable codopants in SrTiO$_3$ mostly act as donor. Our results reveal that despite the response in visible light region, the codoping in TiO$_2$ and SrTiO$_3$ cannot always enhance the photocatalytic activity due to either the formation of recombination centers or the large shift in the conduction band minimum or valence band maximum. Amongst various metal-nonmetal combinations, Mn$_\textrm{Ti}$S$_\textrm{O}$ (i.e. Mn is substituted at Ti site and S is substituted at O site), S$_\textrm{O}$ in anatase TiO$_2$ and Mn$_\textrm{Ti}$S$_\textrm{O}$, Mn$_\textrm{Sr}$N$_\textrm{O}$ in SrTiO$_3$ are the most potent candidates to enhance the photocatalytic efficiency of anatase TiO$_2$ and SrTiO$_3$ under visible light irradiation.
\end{abstract}
\pacs{Valid PACS appear here}% PACS, the Physics and Astronomy
% Classification Scheme.
%\keywords{Suggested keywords}%Use showkeys class option if keyword
%display desired
\maketitle

%\tableofcontents

%%%MAIN TEXT%%%%
\section{Introduction}
Semiconductor-based photocatalysts are seeking the attention ascribed to their potential in utilizing the solar energy to cater to the current energy demand of the world and also, serve the purpose of pollutant degradation~\cite{doi:10.1021/cr1001645,C5TA06983A,B921692H,B800489G,doi:10.1021/cr100454n,Su2013,Takata_2015,C8TA09865D}. Anatase TiO$_2$ and SrTiO$_3$ are two of the metal oxides that can be used in photocatalytic water splitting~\cite{r1, r2, r3, r4, r5, r6, r7, r8, r9, B800489G,doi:10.1098/rsta.2010.0111,C9TA11048H,molecules21070900,Kuo2012,C8CP00844B,C5TA04843E} owing to their suitable band edge positions. However, they could only exploit the UV irradiation of the solar spectrum attributed to their wide band gap of $\sim3.2$ eV. This leads to the low photocatalytic efficiency and limits their application at a commercial level. An efficient photocatalyst should have congenial band gap such that it is wide enough to straddle the redox potential of a desired compound and narrow enough to absorb the visible light of the solar spectrum.  However, despite after significant research endeavors, finding the same has never been easy both experimentally as well as theoretically. Therefore, there is justified interest to reduce the band gap and induce visible light response by means of doping with metals~\cite{kim2004high,liu2013large,wang2014first,C7TA10934B,doi:10.1021/jp049556p,doi:10.1021/jp300910e,B502147B}, nonmetals~\cite{irie2003nitrogen,ohno2003photocatalytic,asahi2001visible,catal_today_2013,C8TA09913H,WANG2004149,Liu_2007,doi:10.1063/1.2403181,C5CP03005F} or their combination~\cite{C3TA01298K,zhu2009band,wang2014band,C2TA00450J,doi:10.1021/ja210610h,doi:10.1021/cm503541u,doi:10.1021/la0353125}. Although the doping can tune the band gap~\cite{r1, r2, r3, r4, r5, r6, r7, r8, r9}, many a time the band edge positions also get changed and the localized deep trap states occur in forbidden region leading to faster recombination. As a consequence, the photocatalytic efficiency gets degraded. Therefore, we need to ensure the suitable band edge positions as well as the passivation of midgap trap states (i.e. recombination centers).

The conduction band minimum (CBm) lies $\sim$ 0.4 eV and 0.8 eV above the reduction potential of water for anatase TiO$_2$~\cite{doi:10.1021/ja954172l} and SrTiO$_3$~\cite{band_edge}, respectively, concomitant with the reduction of water to produce hydrogen under UV irradiation. The transition metal dopant shifts the CBm downwards (towards valence band maximum (VBM)) and hence results in deterioration of reduction power. On the other hand, in case of nonmetal dopant, the band gap reduction takes place by elevating the VBM. However, many a time localized deep trap states appear in forbidden region in both the aforementioned cases. Moreover, the nonmetal doped systems are unstable against exposure to light irradiation~\cite{doi:10.1021/jp902567j}. Therefore, despite the visible light absorption, none of the monodopants (be it metal or nonmetal) are usually suitable for photocatalytic water splitting. This has motivated us to codope the system. The codoping of a metal and a nonmetal is one of the prominent solutions to passivate the trap states and form the shallow impurity states~\cite{doi:10.1021/jp902567j}. Note that for maximum efficiency, the band gap should be $\sim$ 2 eV~\cite{doi:10.1021/cr1002326,C3CP54589J}. With the aid of codoping, the band gap could be tailored such that it induces visible light absorption while retaining the redox powers and thus, enhances the photocatalytic efficiency~\cite{Khaselev425}. %We have already checked that the systems' (N-Mn, S-Mn, and S-Rh codoped SrTiO$_3$) band gap reduction is most favorable from the perspective of their usage in photocatalytic water splitting ($\sim$ 2 eV)~\cite{jpc_sto}. 
Note that the experimental synthesis of individual monodopants (i.e. N, S, Rh and Mn) in TiO$_2$ as well as SrTiO$_3$ have already been done~\cite{Asahi269,doi:10.1021/ja711023z,MATSUMOTO1982419,doi:10.1021/ja403761s,Liu_2007,doi:10.1063/1.2403181,TKACH20055061,doi:10.1080/21663831.2013.856815,doi:10.1021/ja2050315}. A very few experimental studies also exist on N-Mn and S-Mn codoped TiO$_2$ as well~\cite{QUAN2014120,DIVYALAKSHMI2018494}. However, for the codoping (viz. N-Mn, N-Rh, S-Mn and S-Rh) in bulk SrTiO$_3$ and N-Rh, S-Rh in anatase TiO$_2$, any experimental or theoretical reports are hitherto unknown.

In this article, we have studied the role of codopants (N-Mn, N-Rh, S-Mn and S-Rh) in anatase TiO$_2$ and SrTiO$_3$ for enhancing the photocatalytic efficiency. We have shown in Ref.~\cite{scireports2019,jpc_sto} that the substitutional doping is more favorable than the interstitial in TiO$_2$ as well as SrTiO$_3$. Therefore, we have chosen only the substitutional positions (i.e. metal at Ti or Sr site and nonmetal at O site) for codoping. First, we have determined the thermodynamic stability of codoped TiO$_2$ and SrTiO$_3$ under the framework of hybrid density functional theory (DFT)~\cite{PhysRev.136.B864,PhysRev.140.A1133} and \textit{ab initio} thermodynamics at realistic conditions (temperature $(T)$, partial pressure of oxygen $(p_{\textrm{O}_2})$ and doping)~\cite{PhysRevMaterials}. Next, the electronic structure has been analyzed to get insight about the defect states. Further, we have calculated the optical properties to compare the spectral response of all the defected configurations w.r.t pristine counterpart. To explore the best set of codoped combinations for photocatalytic water splitting, the band edge alignment has been carefully carried out. Finally, under codoping, the effective mass has been determined to understand the effect on mobility of positive ions.

\section{Methodology}
We have performed the DFT calculations as implemented in Vienna \textit{ab initio} simulation package (VASP)~\cite{KRESSE199615,PhysRevB.59.1758}. The projector-augmented wave (PAW) pseudopotentials~\cite{PhysRevB.50.17953} have been used to describe the interactions between electrons and ions for all the species. For the energy calculations, hybrid exchange-correlational (xc) functional HSE06~\cite{doi:10.1063/1.2404663} is used. The exact exchange fractions in HSE06 functional used for TiO$_2$ and SrTiO$_3$ are 22\% and 28\%, respectively (see SI ref.~\cite{scireports2019,jpc_sto} for validation of exact exchange fraction). The band gap of 3.15 eV and 3.28 eV are reproduced for TiO$_2$ and SrTiO$_3$ respectively, which are well in agreement with the experimental values~\cite{todorov2017ultrathin,doi:10.1063/1.1415766}. To make the defect to be localized, we have used $2\times2\times1$ (48-atom) and $2\times2\times2$ (40-atom) supercells by replication of TiO$_2$ and SrTiO$_3$ unit cells respectively. The k-grid for Brillouin zone sampling is generated using Monkhorst-Pack~\cite{PhysRevB.13.5188} scheme and all results are checked for convergence w.r.t. the mesh size ($4\times4\times4$). The electronic self-consistency loop for the total energy is converged with a threshold of 0.01 meV. An energy cutoff of 600 eV is used for the plane wave basis set. Note that the spin-polarized calculations have been carried out since the doped systems contain unpaired electrons.

\section{Results}
\subsection{Stability of codoped TiO$_2$ and SrTiO$_3$}
The stability has been determined by calculating the defect formation energy using hybrid DFT and \textit{ab initio} thermodynamics~\cite{Bhattacharya_2014,ekta-jpcc,PhysRevMaterials}. The defect configuration, having minimum formation energy, is the most stable defect. For a defect X with charge state $q$, the formation energy $\textrm{E}_\textrm{f}(\textrm{X}^{q})$ is evaluated as follow~\cite{scireports2019,PhysRevMaterials,PhysRevB.94.094305}:
\begin{equation}\begin{split}
\textrm{E}_\textrm{f}(\textrm{X}^{q}) &= \textrm{E}_\textrm{tot}(\textrm{X}^{q}) - \textrm{E}_\textrm{tot}(\textrm{pristine}^{0}) - \sum_{i} {n}_i\mu_{i}\\
&\quad+ {q}(\mu_\textrm{e} + \textrm{VBM} + \Delta\textrm{V})\textrm{,}
\end{split}\end{equation}
where $\textrm{E}_\textrm{tot}(\textrm{X}^{q})$ and $\textrm{E}_\textrm{tot}(\textrm{pristine}^{0})$ are the energies of defected supercell (charged) and pristine supercell (neutral), respectively, calculated using hybrid DFT. ${n}_i$ is the number of species $i$ added to or removed from the pristine supercell and $\mu_{i}$'s are the corresponding chemical potentials, which is selected with reference to the total energy ($\textrm{E}_\textrm{tot}({i}^{0})$) of species $i$. Therefore, $\mu_{i} = \Delta\mu_{i} + \textrm{E}_\textrm{tot}({i})$ ($i$ = N, S, O, Mn, Rh, Sr, or Ti), where $\Delta\mu_{i}$'s are chosen according to the environmental growth conditions. $\mu_\textrm{e}$ is the chemical potential of the electron, which is the energy required to exchange electrons between the system and the electrons' reservoir. It is varied from VBM to CBm of the pristine supercell. $\Delta\textrm{V}$ is the core level alignment between pristine neutral and defected supercell.

The chemical potential of a species incorporates the effect of temperature and pressure. For oxygen, the $\Delta\mu_\textrm{O}$ as a function of temperature ($T$) and the partial pressure of oxygen ($p_{\textrm{O}_2}$) is calculated as follow~\cite{Bhattacharya_2014}:
\begin{equation}\begin{split}
\Delta\mu_\textrm{O}(T, p_{\textrm{O}_2}) &= \frac{1}{2}\left[ -k_\textrm{B}T \ln\left[\left(\frac{2\pi m}{h^2}\right)^\frac{3}{2}\left(k_\textrm{B}T\right)^\frac{5}{2}\right]\right.
\\
&\quad+ k_\textrm{B}T \ln p_{\textrm{O}_2} - k_\textrm{B}T \ln \left(\frac{8\pi^2I_Ak_\textrm{B}T}{h^2}\right)
\\
&\quad+ k_\textrm{B}T \ln \left[1-\exp\left(\frac{-h\nu_\textrm{OO}}{k_\textrm{B}T}\right)\right]
\\
&\quad\left.-  k_\textrm{B}T \ln \mathcal{M} + k_\textrm{B}T \ln \sigma \right]\textrm{,}
\end{split}\end{equation}
where $m$ is the mass, $I_A$ is the moment of inertia of $\textrm{O}_2$ molecule, $\nu_\textrm{OO}$ is the O-O stretching frequency, $\mathcal{M}$ is the spin multiplicity and $\sigma$ is the symmetry number.
\begin{figure*}
	\centering
	\includegraphics[width=0.9\textwidth]{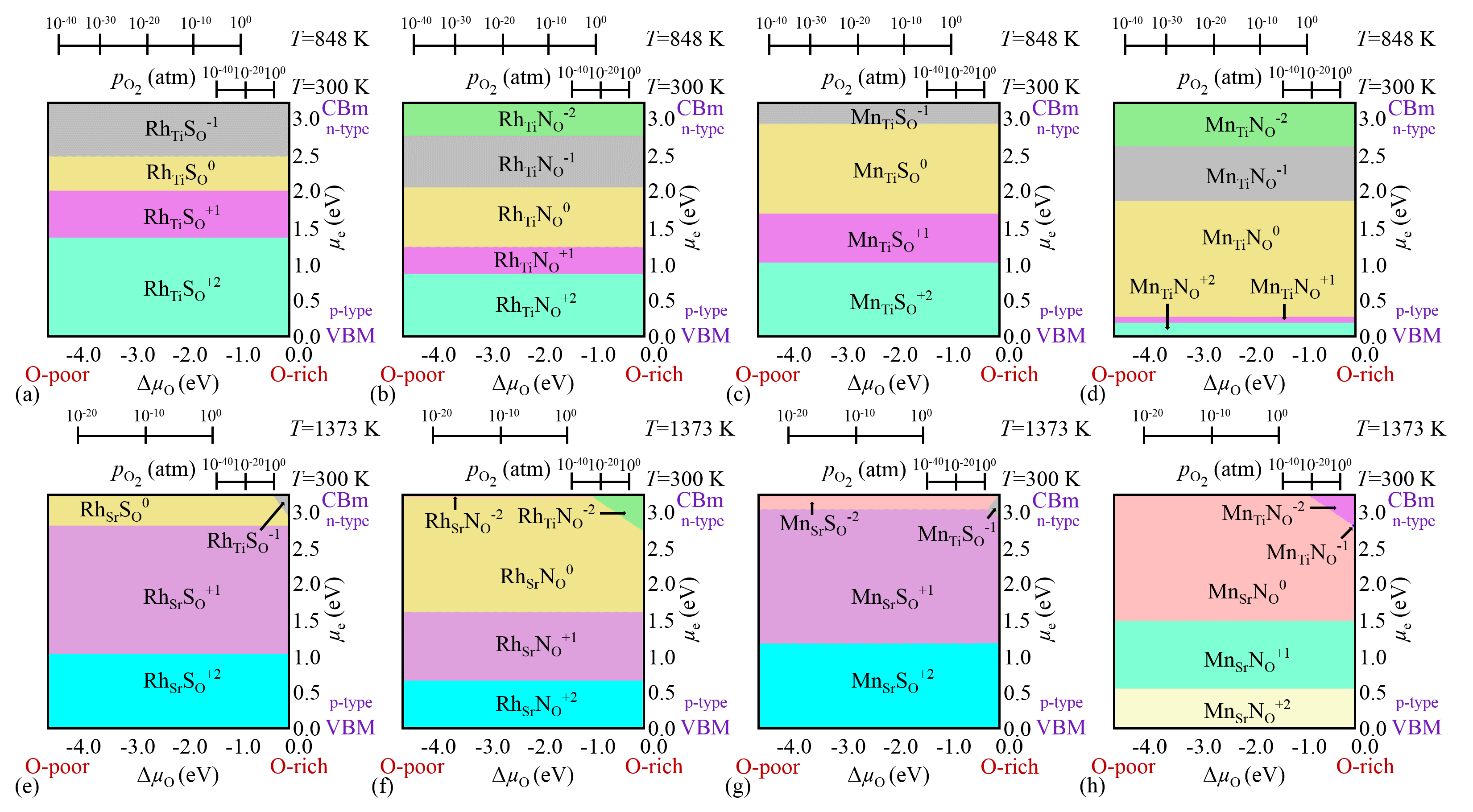} 
	\caption{3D phase diagram that shows the most stable phases of (a) Rh$_\textrm{Ti}$S$_\textrm{O}$, (b) Rh$_\textrm{Ti}$N$_\textrm{O}$, (c) Mn$_\textrm{Ti}$S$_\textrm{O}$ and (d) Mn$_\textrm{Ti}$N$_\textrm{O}$ codoped TiO$_2$ charged configurations, and (e) S-Rh, (f) N-Rh, (g) S-Mn and (h) N-Mn~\cite{jpc_sto} codoped SrTiO$_3$ charged configurations with minimum formation energy as a function of $\Delta\mu_\textrm{O}$ and $\mu_\textrm{e}$. Here, $\Delta\mu_\textrm{O}$ (on x-axis) is varied according to environmental growth conditions ($T$ and $p_{\textrm{O}_2}$ (on top axes)) and $\mu_\textrm{e}$ (on y-axis) is varied from VBM to CBm of pristine system. The experimental growth conditions for TiO$_2$ are $T$ = 848 K, $p_{\textrm{O}_2} = 2.6\times10^{-8}$ atm~\cite{PhysRevB.79.085401} and for SrTiO$_3$ are $T$ = 1373 K, $p_{\textrm{O}_2} = 1$ atm~\cite{WANG2004149}.}
	\label{fig_def}
\end{figure*}

For TiO$_2$ and SrTiO$_3$, the equilibrium growth conditions are:
\begin{equation}\begin{split}
\Delta\mu_\textrm{Ti} + 2\Delta\mu_\textrm{O} = \Delta\textrm{H}_\textrm{f}(\textrm{TiO}_2)\textrm{,}
\\
\Delta\mu_\textrm{Sr} + \Delta\mu_\textrm{Ti} + 3\Delta\mu_\textrm{O} = \Delta\textrm{H}_\textrm{f}(\textrm{SrTiO}_3)\textrm{,}
\end{split}\label{mu1}\end{equation}
respectively. Here, $\Delta\textrm{H}_\textrm{f}$ denotes the enthalpy of formation. The bounds are imposed on chemical potentials to restrain the formation of secondary phases (viz. TiO$_2$, SrO in SrTiO$_3$) as follow:
\begin{equation}\begin{split}
\Delta\mu_\textrm{Ti} + 2\Delta\mu_\textrm{O} \leq \Delta\textrm{H}_\textrm{f}(\textrm{TiO}_2)\textrm{,}
\\
\Delta\mu_\textrm{Sr} + \Delta\mu_\textrm{O} \leq \Delta\textrm{H}_\textrm{f}(\textrm{SrO})\textrm{,}
\\
\Delta\mu_{i} \leq 0\textrm{.}
\end{split}\label{mu2}\end{equation}
The chemical potentials are determined by imposing bounds on the formation of the precursors (MnO$_2$, TiO$_2$, TiS$_2$, Rh$_2$O$_3$), and the aforementioned Eqs.~\ref{mu1} and~\ref{mu2}.

In our previous findings, we have seen that the substitutional defect is more favorable in comparison to interstital and also, observed that the monodopants in general are not suitable for photocatalytic application~\cite{scireports2019,jpc_sto}. Therefore, we have considered here the codoped cases (metal (Rh or Mn) substituted at Ti or Sr site, and nonmetal (N or S) substituted at O site). Fig.~\ref{fig_def} shows the stable phases of codoped TiO$_2$ and SrTiO$_3$ with minimum formation energy. These stable phases include charged defects in addition to neutral defects due to the uncompensated charge.  The positive charge states are stable near VBM (in p-type host), whereas the negative charge states are stable near CBm (in n-type host). The Rh$_\textrm{Ti}$S$_\textrm{O}$ and Mn$_\textrm{Ti}$S$_\textrm{O}$ codoped TiO$_2$ could be stable in +2, +1, 0, and -1 charge states [see Figs.~\ref{fig_def}(a) and~\ref{fig_def}(c)]. These codopants will act as donor in p-type host (near VBM), and acceptor in n-type host (near CBm). The charge state -2 is not stable in both the cases. As these transition metals have partially filled d-orbitals, it could accept electrons from the host as well as donate electrons to the host. Figs.~\ref{fig_def}(b) and~\ref{fig_def}(d) show that Rh$_\textrm{Ti}$N$_\textrm{O}$ and Mn$_\textrm{Ti}$N$_\textrm{O}$ codoped TiO$_2$ will be stable in -2 charge state in addition to +2, +1, 0 and -1, since N has one electron less than O, it could accept one electron extra in comparison to S (S has same number of valence electrons as for O). In Mn$_\textrm{Ti}$S$_\textrm{O}$ and Mn$_\textrm{Ti}$N$_\textrm{O}$ codoped TiO$_2$, for a large range of $\mu_\textrm{e}$, neutral charge state is more favorable. The Sr site is more favorable than Ti site for substitution in SrTiO$_3$ [see Figs.~\ref{fig_def}(e-h)]. The Ti site could be substituted in O-rich (Ti-poor) conditions near CBm (in n-type host). Mostly, Rh$_\textrm{Sr}$S$_\textrm{O}$, Rh$_\textrm{Sr}$N$_\textrm{O}$, Mn$_\textrm{Sr}$S$_\textrm{O}$, and Mn$_\textrm{Sr}$N$_\textrm{O}$ codopants act as donor as they are stable in +2 and +1 charge states for a wide range of $\mu_\textrm{e}$ or in neutral charge state. However, when metal is substituted at the Ti position, the defect configuration will act as an acceptor. Similar to the codoped TiO$_2$, the Rh$_\textrm{Ti}$N$_\textrm{O}$, and Mn$_\textrm{Ti}$N$_\textrm{O}$ codoped SrTiO$_3$ will get stabilized after accepting one extra electron in comparison to Rh$_\textrm{Ti}$S$_\textrm{O}$, and Mn$_\textrm{Ti}$S$_\textrm{O}$.
\subsection{Electronic density of states (DOS)}
\begin{figure*}
	\centering
	\includegraphics[width=0.8\textwidth]{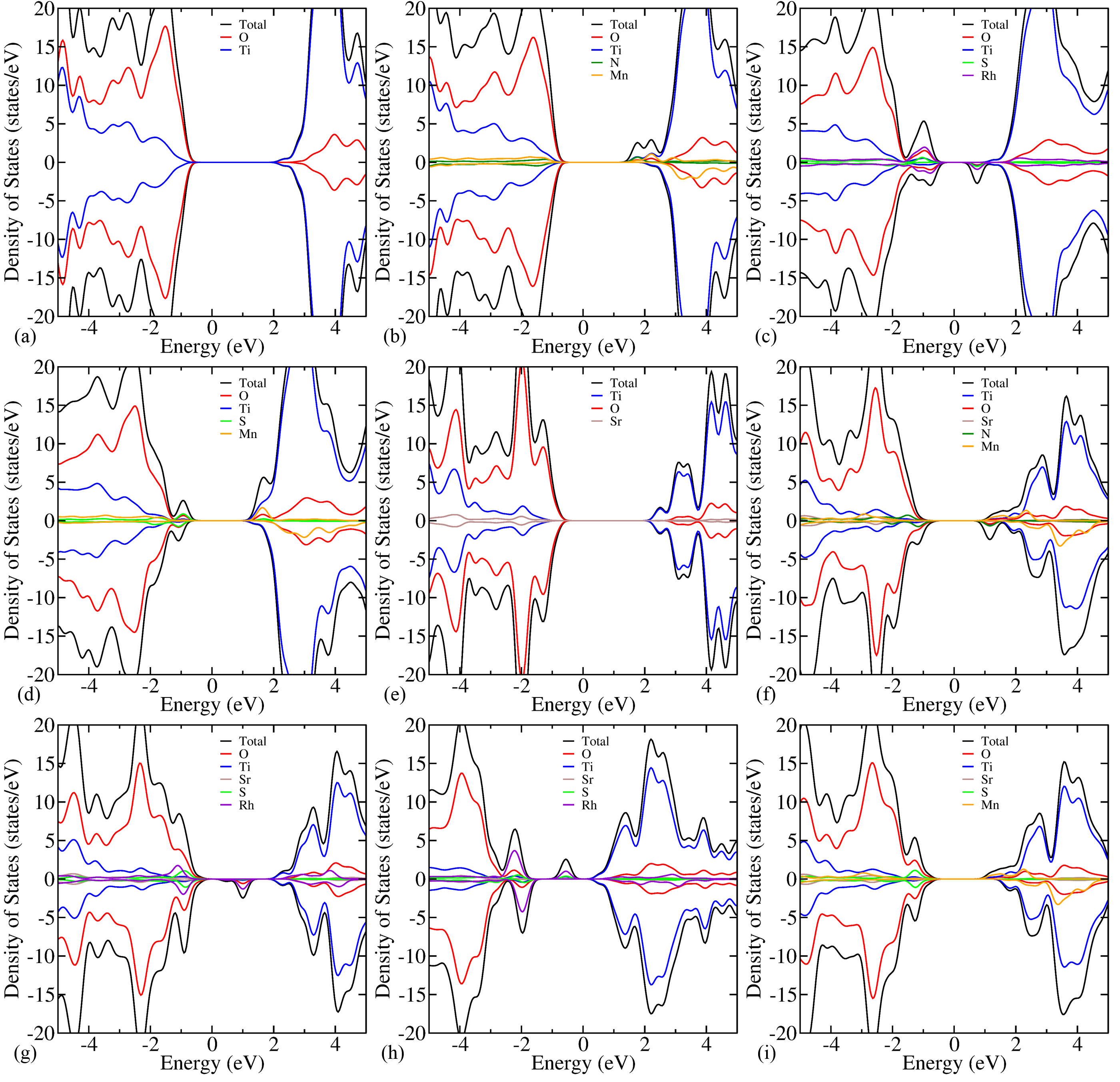}
	\caption{Atom projected density of states of (a) pristine TiO$_2$, (b) Mn$_\textrm{Ti}$N$_\textrm{O}$, (c) Rh$_\textrm{Ti}$S$_\textrm{O}$, (d) Mn$_\textrm{Ti}$S$_\textrm{O}$ codoped TiO$_2$, and (e) pristine SrTiO$_3$, (f) Mn$_\textrm{Ti}\textrm{N}_\textrm{O}$~\cite{jpc_sto}, (g) Rh$_\textrm{Ti}\textrm{S}_\textrm{O}$, (h) Rh$_\textrm{Sr}$S$_\textrm{O}$, and (i) Mn$_\textrm{Ti}$S$_\textrm{O}$ codoped SrTiO$_3$.}
	\label{fig_DOS}
\end{figure*}
The defect states could be seen by means of electronic density of states. Fig.~\ref{fig_DOS} shows the atom projected density of states for pristine and codoped TiO$_2$ as well as SrTiO$_3$. In the pristine systems, near Fermi level, the valence band is contributed by O 2p orbitals and the conduction band is contributed by Ti 3d orbitals [see Figs.~\ref{fig_DOS}(a) and~\ref{fig_DOS}(e)]. The DOS is symmetric w.r.t. the spin alignments ascribed to the paired electrons in the system. On doping, the DOS becomes asymmetric attributable to unpaired electrons. The deep trap states arise in N$_\textrm{O}$, Rh$_\textrm{Ti}$ and Rh$_\textrm{Ti}$N$_\textrm{O}$ doped TiO$_2$, that increase the recombination rate and deteriorate the photocatalytic efficiency [see Figs. S1(a), S1(d) and S1(e)]. There is very slight reduction in band gap for Mn$_\textrm{Ti}$ doped TiO$_2$. Hence, it's not inducing the effective visible light absorption [see Fig. S1(c)]. In Mn$_\textrm{Ti}$N$_\textrm{O}$ codoped TiO$_2$, the N orbitals and Mn orbitals shift down the CBm, leading to the reduction in band gap and induce the visible light absorption [see Fig.~\ref{fig_DOS}(b)]. However, this shift will lower down the CBm and thus, not efficient for reduction of water to produce hydrogen. The VBM is elevated in Rh$_\textrm{Ti}$S$_\textrm{O}$ codoped TiO$_2$, which is caused by the S and Rh orbitals contribution [see Fig.~\ref{fig_DOS}(c)]. Furthermore, the CBm is also shifted down due to the unoccupied states of S and Rh orbitals. Therefore, despite its spectral response in visible region, it cannot be used for producing oxygen via water reduction ascribed to the large shift in VBM. The band gap reduction is induced by the S orbitals in S$_\textrm{O}$ doped TiO$_2$ as the S orbitals energies lie higher than the N orbitals [see Fig. S1(b)]. Also, in Mn$_\textrm{Ti}$S$_\textrm{O}$ codoped TiO$_2$, the S orbitals elevate the VBM as they have higher energy than the O orbitals and the Mn orbitals contribute to CBm [see Fig.~\ref{fig_DOS}(d)]. This will enhance the photocatalytic efficiency, as the band gap becomes 2.2 eV in both the aforementioned cases and the band edges straddle the redox potentials of water.

For SrTiO$_3$, N$_\textrm{O}$ and S$_\textrm{O}$ behave similar to the case of N$_\textrm{O}$ and S$_\textrm{O}$ monodoped TiO$_2$ [see Figs. S1(f) and S1(g)]. However, the reduction in band gap for S$_\textrm{O}$ is small in comparison to S$_\textrm{O}$ doped TiO$_2$. S$_\textrm{O}$ doped SrTiO$_3$ has the band gap of 2.59 eV and hence, responses to visible light irradiation. For Mn$_\textrm{Ti}$ monodoped SrTiO$_3$, the Mn orbital contributes to the CBm and lowers down it [see Fig. S1(h)]. Therefore, its spectral response expands to visible light irradiation (band gap is 2.57 eV). In Rh$_\textrm{Ti}$ monodoped SrTiO$_3$, the unoccupied states of Rh orbitals appear at VBM, and the difference between highest occupied and lowest unoccupied state is 0.23 eV [see Fig. S1(i)]. Thus, it is not a promising candidate for enhanced photocatalytic activity. The lowering of CBm in Rh$_\textrm{Sr}$ is occurred due to the Rh localized states contribution to CBm and therefore, it doesn't have enough reduction power to produce hydrogen via water splitting [see Fig. S1(j)]. There is no reduction in band gap for Mn$_\textrm{Sr}$ monodoped SrTiO$_3$, as the Mn orbitals contribute deep inside the valence and conduction band [see Fig. S1(k)].  
Likewise Mn$_\textrm{Ti}$N$_\textrm{O}$ codoped TiO$_2$, the reduction in band gap of Mn$_\textrm{Ti}$N$_\textrm{O}$ codoped SrTiO$_3$ is occurred by lowering of the CBm as well as elevation of the VBM [see Fig.~\ref{fig_DOS}(f)]. However, the shift in CBm is large enough such that its reduction power is deteriorated. In Rh$_\textrm{Ti}$S$_\textrm{O}$ and Rh$_\textrm{Sr}$S$_\textrm{O}$ codoped SrTiO$_3$, the deep trap states arise in the forbidden region, that increase the recombination of photogenerated charge carriers and thus, degrade the photocatalytic activity [see Figs.~\ref{fig_DOS}(g) and~\ref{fig_DOS}(h)]. Since there is occurrence of trap states in Rh$_\textrm{Ti}$N$_\textrm{O}$ and Mn$_\textrm{Sr}$S$_\textrm{O}$ codoped SrTiO$_3$, it will result in poor photocatalytic activity [see Figs. S1(l) and S1(o)]. The Rh and N orbitals states elevate the VBm of Rh$_\textrm{Sr}$N$_\textrm{O}$ codoped SrTiO$_3$, which results in a band gap of 2.69 eV [see Fig. S1(m)]. The Mn$_\textrm{Ti}$S$_\textrm{O}$ codoped SrTiO$_3$ has the band gap of 1.95 eV. The VBM elevation is concomitant with occurrence of S orbitals at VBM and the Mn orbitals at CBm [see Fig.~\ref{fig_DOS}(i)]. The shifts in CBm and VBM are such that the band edges straddle the redox potential levels of water. Similarly, for Mn$_\textrm{Sr}$N$_\textrm{O}$ codoped SrTiO$_3$, the band gap is 2.17 eV, and the defect states are shallower, which serves the purpose of efficient photocatalyst [see Fig. S1(n)].
\subsection{Optical properties}
\begin{figure}[h]
	\centering
	\includegraphics[width=0.48\textwidth]{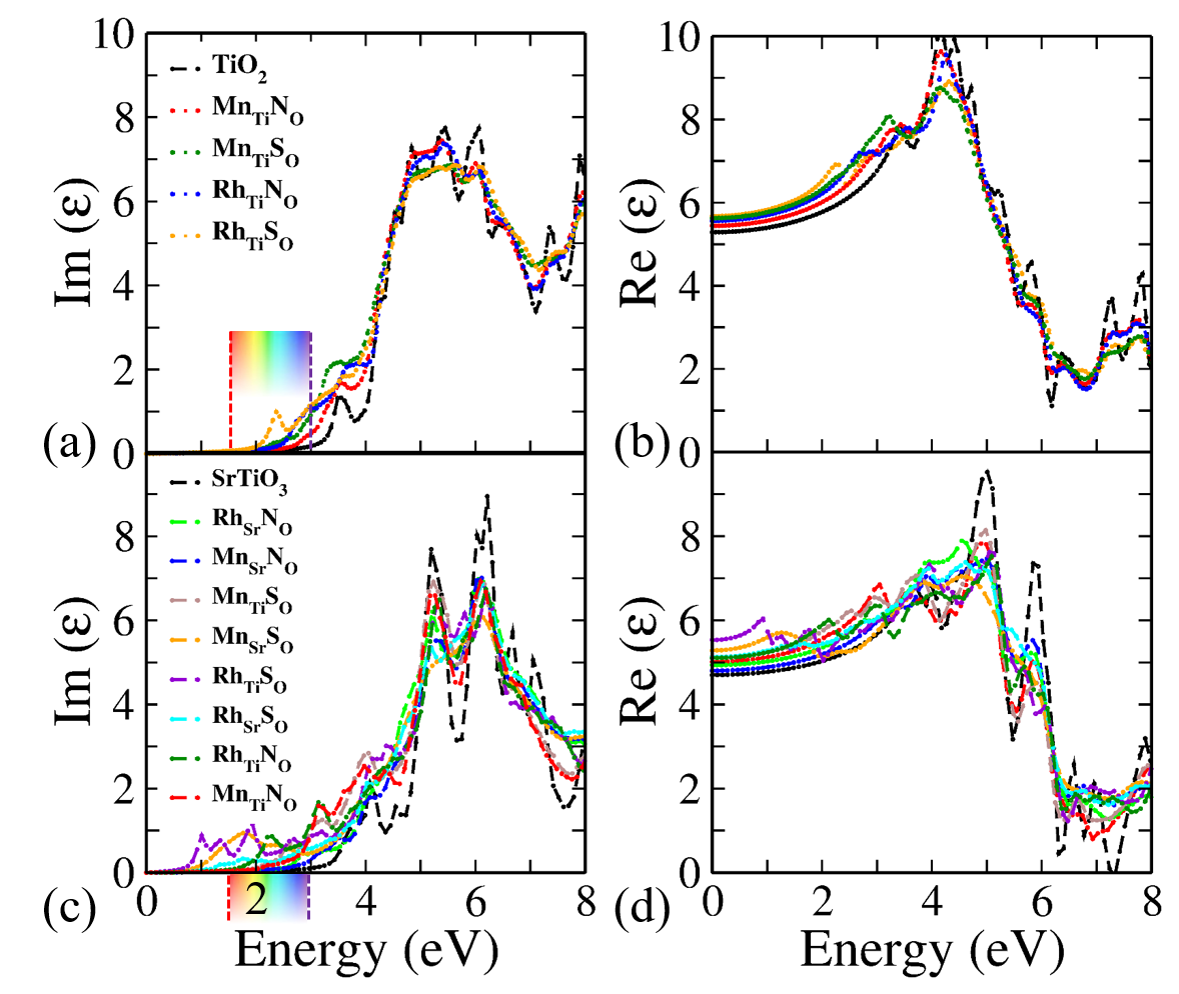}
	\caption{Spatially average (a) imaginary (Im $\varepsilon$) and (b) real (Re $\varepsilon$) part of the dielectric function for (un)doped TiO$_2$, (c) imaginary (Im $\varepsilon$) and (d) real (Re $\varepsilon$) part for (un)doped SrTiO$_3$.}
	\label{fig_opt}
\end{figure}

The optical spectra has been determined by calculating the frequency dependent complex dielectric function $\varepsilon(\omega) = \textrm{Re} (\varepsilon) + \textrm{Im}(\varepsilon)$ using HSE06 functional. The real part $\textrm{Re}(\varepsilon)$ and the imaginary part $\textrm{Im}(\varepsilon)$ are associated with the electronic polarizability and optical absorption of the material, respectively. The sum of all possible transitions from the occupied to the unoccupied states gives the direct interband transition, which is reflected in the imaginary part of the dielectric function. The imaginary and real part for codoped anatase TiO$_2$ and SrTiO$_3$ are shown in Fig.~\ref{fig_opt} (the results for monodoped TiO$_2$ and SrTiO$_3$ are shown in Fig. S2). Note that anatase TiO$_2$ has tetragonal structure. Therefore, the optical anisotropy is also associated with it. The detailed discussion of optical anisotropy is already done in our previous work~\cite{C9TC05002G}.
Therefore, here we have shown only the averaged (x, y, z polarizations) for  imaginary and real part of the dielectric function. The imaginary part of dielectric function shows the first peak at 3.56 eV for pristine TiO$_2$ as shown in Fig.~\ref{fig_opt}(a) [matching with the previous works, which is 3.8 eV~\cite{PhysRevB.82.045207}].  The peaks are shifted to lower energy for codoped cases. This enhances the visible light absorption of anatase TiO$_2$. The static real part of the dielectric function (at $\omega = 0$) for TiO$_2$ is found to be  5.28 [see Fig.~\ref{fig_opt}(b)], which is very close to the experimental value i.e., 5.62~\cite{doi:10.1063/1.435102}.  On codoping its value is increased.

For the case of cubic SrTiO$_3$, the spatially average imaginary and real part of dielectric function are shown in Figs.~\ref{fig_opt}(c) and~\ref{fig_opt}(d), respectively.  The static real part of the dielectric function for pristine SrTiO$_3$ is estimated as 4.7 (experimental value is  5.27~\cite{PhysRev.140.A651}) and its value is increased with codopants [see Fig.~\ref{fig_opt}(d)]. The first absorption peak is observed at 4.08 eV for pristine SrTiO$_3$ as shown in Fig.~\ref{fig_opt}(c) [experimental value is 4.7 eV~\cite{PhysRev.140.A651}].  Likewise in anatase TiO$_2$, the peaks are shifted to visible region for the codoped cases. Note that the optical properties in the high energy range are controlled by the electronic transitions between O 2p states and Ti 3d states. Therefore, the spectra of all the configurations are nearly identical in high energy range. However, the optical properties in low energy range (less than 3 eV) are different, these are affected by the transitions involving the impurity states. The observed visible light absorption could be ascribed to the presence of the dopant states (as shown in DOS near Fermi-level), which reduce the electron transition gap for optical absorption. This leads to a new absorption edge in the visible light region.
\subsection{Band edge alignment}
\begin{figure}[h]
	\centering
	\includegraphics[width=0.48\textwidth]{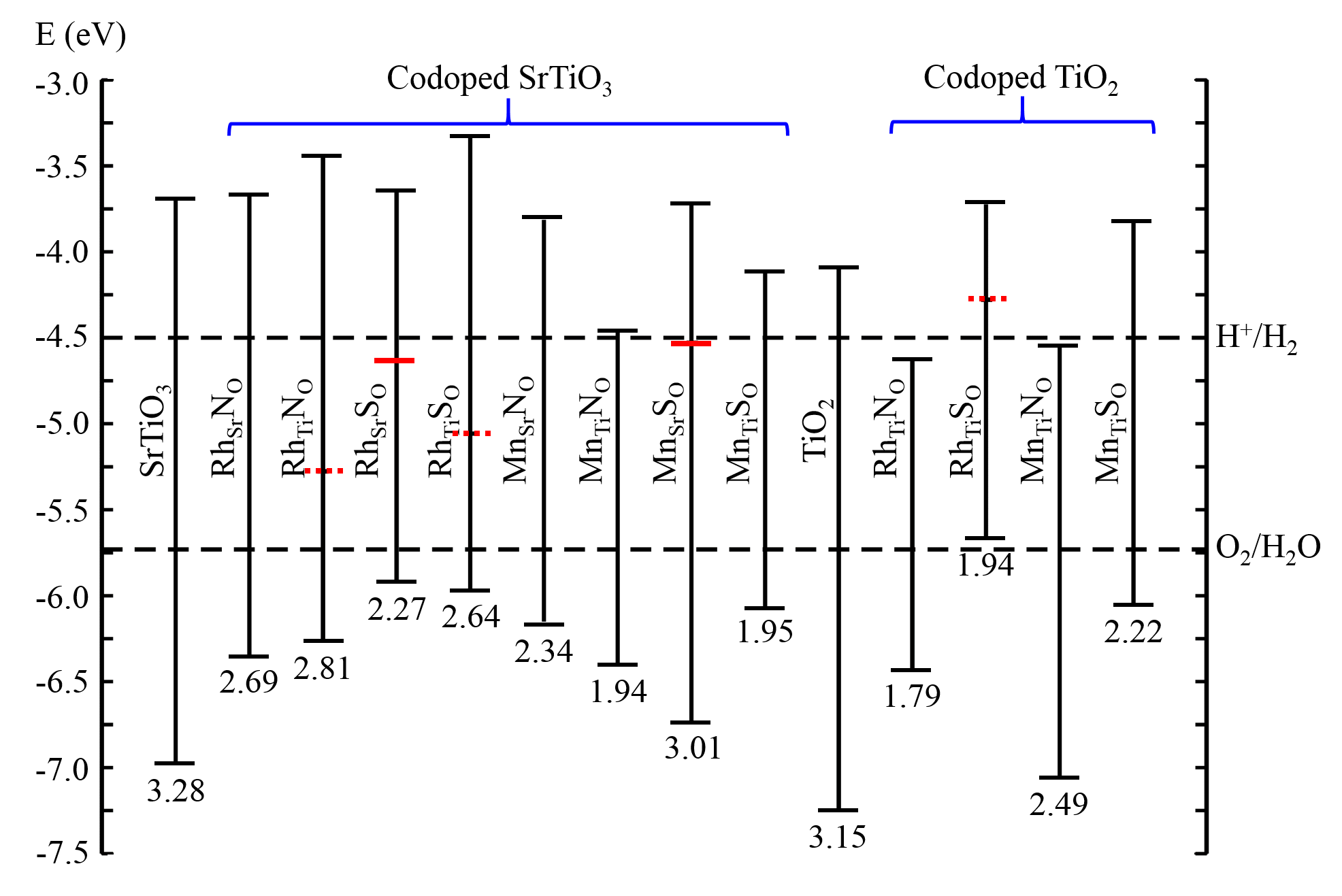}
	\caption{Band edge alignment of (un)doped SrTiO$_3$ and TiO$_2$ w.r.t. water reduction and oxidation potential levels (H$^+$/H$_2$, O$_2$/H$_2$O). The solid and dashed red line in forbidden region are representing the highest occupied and lowest unoccupied states, respectively.}
	\label{fig_band_edge_align}
\end{figure}

The band edge alignment has been performed to obtain the potential candidates for photocatalytic water splitting. The CBm should lie above water reduction potential and VBM should lie below water oxidation potential for overall water splitting. First we align the band edges of pristine TiO$_2$ and SrTiO$_3$ w.r.t. water redox potential~\cite{doi:10.1021/ja954172l,band_edge} and further, we align the band edges of defected configurations by observing the shift in CBm and VBM w.r.t. the pristine system [see Fig.~\ref{fig_band_edge_align}]. The band edge alignment for monodoped anatase TiO$_2$ and SrTiO$_3$ are shown in Fig. S3. The monodoped N$_\textrm{O}$ is not suitable in both the cases (anatase TiO$_2$ and SrTiO$_3$), as it results in deep trap states. The latter increases the recombination and decreases the mobility of photogenerated charge carriers [see Fig. S3]. Likewise, for Rh dopant (in monodoping as well as in codoping), there is occurrence of trap states. These states degrade the photocatalytic efficiency. Therefore, mono- and codopoing of Rh with a nonmetal could reduce the band gap, but it cannot be an efficient photocatalyst in TiO$_2$ as well as in SrTiO$_3$.

The monodoped S$_\textrm{O}$ in both anatase TiO$_2$ as well as SrTiO$_3$ could enhance the photocatalytic efficiency and split water as their band edges straddle the redox potential of water [see Fig. S3]. However, in S$_\textrm{O}$ monodoped SrTiO$_3$, the band gap (2.59) is slightly higher than the desirable band gap ($\sim$ 2 eV~\cite{doi:10.1021/cr1002326,C3CP54589J}), and thus, its efficiency will be smaller. Similarly, for Mn$_\textrm{Ti}$ monodoped SrTiO$_3$, the band gap is 2.57 eV, and due to shift of its CBm towards Fermi level, its reduction power will be degraded [see Fig. S3]. On the other hand, for Mn$_\textrm{Sr}$ monodoped SrTiO$_3$, and Mn$_\textrm{Ti}$ monodoped anatase TiO$_2$, the slight change in band gap are observed and thus, these can not enhance the photocatalytic activity.
In Rh$_\textrm{Ti}$S$_\textrm{O}$ codoped TiO$_2$, since there is manifestation of deep unoccupied states as well as the VBM lies above the oxidation potential of water, it could not be utilized for photocatalytic overall water splitting [see Fig.~\ref{fig_band_edge_align}]. For Rh$_\textrm{Ti}$N$_\textrm{O}$ and Mn$_\textrm{Ti}$N$_\textrm{O}$ codoped TiO$_2$, the CBm lies below the reduction potential of water, and thus, cannot produce hydrogen via water splitting. In TiO$_2$, only the Mn$_\textrm{Ti}$S$_\textrm{O}$ codoping is the potential candidate for overall photocatalytic water splitting as it has a desirable band gap of 2.22 eV and it does not contain the trap states in forbidden region while retaining the sufficient reduction and oxidation power for hydrogen evolution reaction (HER) as well as oxygen evolution reaction (OER). Similarly, in SrTiO$_3$, except Rh$_\textrm{Sr}$N$_\textrm{O}$, the Rh doping does not aid in enhancing the photocatalytic activity ascribed to the formation of recombination centers. Rh$_\textrm{Sr}$N$_\textrm{O}$ defect configuration enhances the photocatalytic efficiency, however its band gap (2.69 eV) is little bit larger in comparison to the maximum efficient photocatalyst ($\sim$ 2 eV). In Mn$_\textrm{Sr}$S$_\textrm{O}$, since  the occupied deep states lie below the CBm and also these are not the shallow impurity levels, this configuration is not a desirable photocatalyst. The reduction in band gap for Mn$_\textrm{Ti}$N$_\textrm{O}$ is concomitant with the lowering of CBm, that deteriorates its reduction power. The Mn$_\textrm{Sr}$N$_\textrm{O}$, and Mn$_\textrm{Ti}$S$_\textrm{O}$ codoped SrTiO$_3$ configurations are the potential candidates for overall photocatalytic water splitting attributable to their desirable band gap ($\sim$ 2 eV) with congenial band edge positions.
\subsection{Band structure and effective mass}
To see the effect on mobility due to the defects, we have calculated the effective mass of charge carriers (using HSE06) of those systems, which could be promising candidates for overall photocatalytic water splitting (see Table~\ref{tbl:example}).
\begin{table}[h]
	\small
	\caption{\ Effective masses (in terms of free-electron mass m$_\textrm{e}$) at the band edges. The masses $\textrm{m}_\textrm{he}, \textrm{m}_\textrm{le},\textrm{m}_\textrm{hh}, \textrm{and}\;\textrm{m}_\textrm{lh}$ correspond to heavy-electron, light-electron, heavy-hole, and light-hole band, respectively.}
	\label{tbl:example}
	\begin{tabular*}{0.48\textwidth}{@{\extracolsep{\fill}}lllll}
		\hline
		Systems & $\textrm{m}_\textrm{he}$ & $\textrm{m}_\textrm{le}$ & $\textrm{m}_\textrm{hh}$ & $\textrm{m}_\textrm{lh}$\\
		\hline
		pristine SrTiO$_3$ & $5.18$ & $0.38$ & $-10.36$ & $-0.74$\\
		Mn$_\textrm{Sr}$N$_\textrm{O}$ codoped SrTiO$_3$ & $3.04$ & -- & -- & $-1.53$\\
		Mn$_\textrm{Ti}$S$_\textrm{O}$ codoped SrTiO$_3$ & -- & $0.25$ & -- & $-0.66$\\
		pristine TiO$_2$ & -- & $0.39$ & $-1.57$ & -- \\
		Mn$_\textrm{Ti}$S$_\textrm{O}$ codoped TiO$_2$ & -- & $0.45$ & $-9.23$ & -- \\
		S$_\textrm{O}$ monodoped TiO$_2$ & -- & $0.41$ & $-2.84$ & -- \\
		\hline
	\end{tabular*}
\end{table}
\begin{figure}[h]
	\centering
	\includegraphics[width=0.48\textwidth]{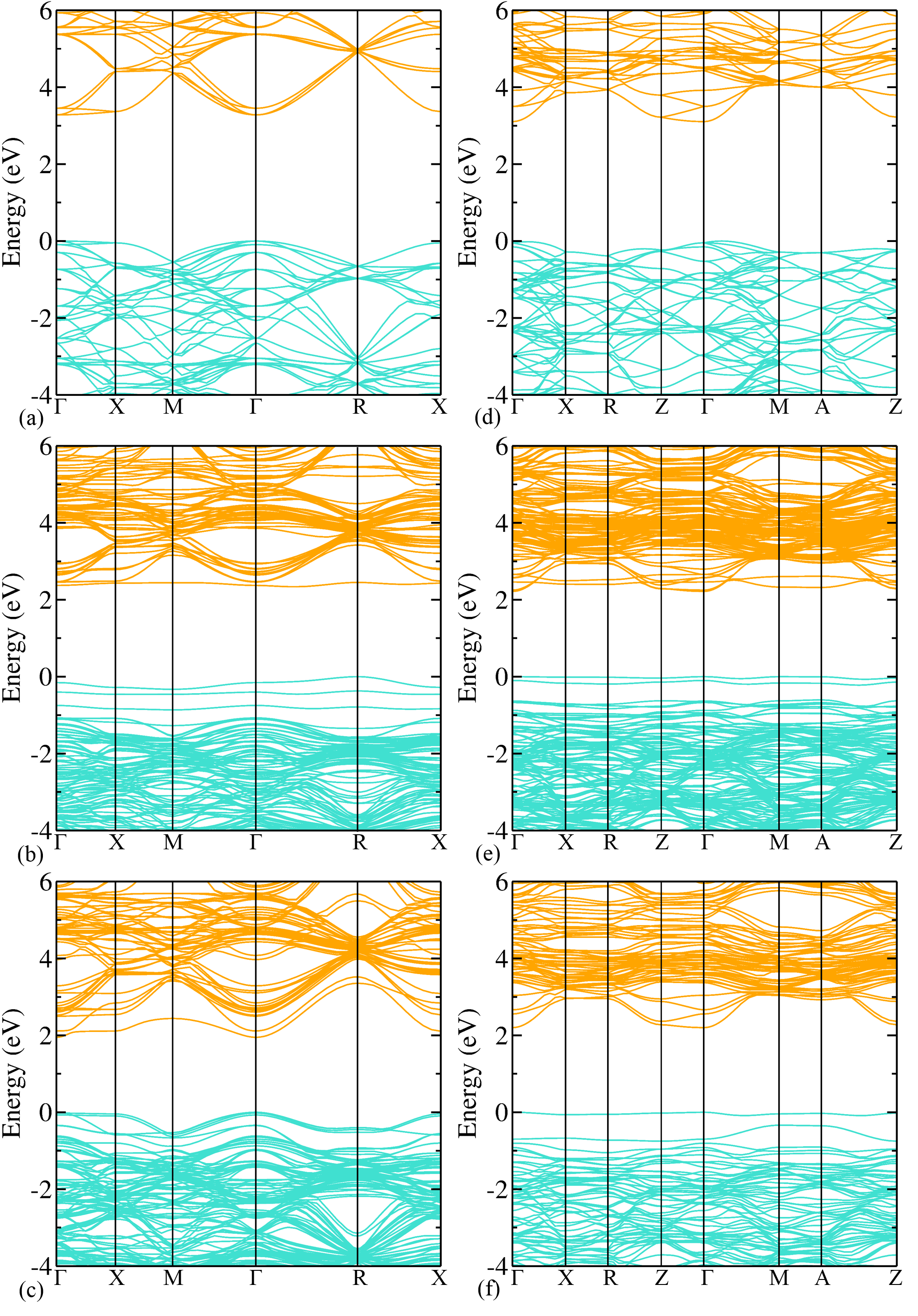}
	\caption{Band structure calculated using hybrid (HSE06) functional of (a) pristine, (b) Mn$_\textrm{Sr}$N$_\textrm{O}$ codoped~\cite{jpc_sto}, (c) Mn$_\textrm{Ti}$S$_\textrm{O}$ codoped SrTiO$_3$ and (d) pristine, (e) Mn$_\textrm{Ti}$S$_\textrm{O}$ codoped, (f) S$_\textrm{O}$ monodoped TiO$_2$.}
	\label{fig_band_str}
\end{figure}
These are obtained from the relation of effective mass ($m^*$) with second derivative of energy with respect to $k$ (wave vector) at the band edges:
\begin{equation}
\frac{1}{m^*} = \frac{1}{\hbar^2}\frac{d^2E}{dk^2}\textrm{,}
\end{equation}
where $\hbar$ is the reduced Planck constant. The effective masses of charge carriers of pristine SrTiO$_3$ are validated with Ref.~\cite{PhysRevB.84.201304,doi:10.1063/1.1570922,C8TA09022J}. Except for the heavy hole, all are matching well. For pristine SrTiO$_3$, the effective masses are calculated along $\Gamma$-X high symmetry path. Pristine has degenerate bands at the $\Gamma$ k-point [see Fig.~\ref{fig_band_str}(a)].  In contrast to pristine SrTiO$_3$, rest of the cases have non-degenerate bands (highest occupied and lowest unoccupied) [see Fig.~\ref{fig_band_str}]. The electron's effective mass of Mn$_\textrm{Sr}$N$_\textrm{O}$ codoped SrTiO$_3$ is 3.04m$_\textrm{e}$ and 5.09m$_\textrm{e}$ along CBm-X and CBm-$\Gamma$ path, respectively, and the hole's effective mass is -1.53m$_\textrm{e}$ and -2.58m$_\textrm{e}$ along R-X and R-$\Gamma$, respectively. These different values along different directions indicate the anisotropic nature of effective mass. For Mn$_\textrm{Ti}$S$_\textrm{O}$ codoped SrTiO$_3$, the effective mass of both the charge carriers (calculated along $\Gamma$-X direction) is decreased. It is also clear from the large curvature of the bands around CBm and VBM in comparison to pristine SrTiO$_3$ [see Fig.~\ref{fig_band_str}(c)].

In pristine TiO$_2$ the CBm is at $\Gamma$ k-point and there is no degeneracy [see Fig.~\ref{fig_band_str}(d)]. The electron's effective mass is 0.39m$_\textrm{e}$ along $\Gamma$-Z high-symmetry path, and the effective mass of hole is -1.57m$_\textrm{e}$ along VBM-Z and -1.61m$_\textrm{e}$ along VBM-$\Gamma$ direction. For Mn$_\textrm{Ti}$S$_\textrm{O}$ codoped TiO$_2$ and S$_\textrm{O}$ monodoped TiO$_2$, the electron's effective mass (along $\Gamma$-Z) is comparable with pristine, whereas the hole's effective mass (along VBM-Z) is increased. These increments are also evident from the smaller curvature of the bands around the band edges [see Figs.~\ref{fig_band_str}(e) and~\ref{fig_band_str}(f)]. For larger effective mass, the mobility will be smaller and the recombination rate will also be greater. Therefore, from Table~\ref{tbl:example}, we can see that in case of Mn$_\textrm{Ti}$S$_\textrm{O}$ codoped SrTiO$_3$, the mobility of charge carriers will be large, and for rest of the cases, the effective mass values are comparable and the mobility will not be affected much. This is because, the mobility depends on both the effective mass and scattering (relaxation) time. On doping, the scattering rate is expected to get decreased as the degeneracy will be lifted. As a consequence of this, despite of small increment in effective mass, the mobility will not be affected very much especially here the doping concentration is low. These effective mass studies should assist future experimental as well as theoretical investigations to tailor the transport properties of the system.

\section{Conclusions}
In summary, we have evaluated the thermodynamic stability of (un)doped anatase TiO$_2$ and SrTiO$_3$ using hybrid DFT and \textit{ab initio} thermodynamics. We have found that the codopants in TiO$_2$ could act as donor (in p-type host) as well as acceptor (in n-type host). However, the most stable codopants (codoping of metal at Sr site and nonmetal at O site) in SrTiO$_3$ mostly act as donors. The codoping expands the spectral response and induce visible light in both the cases. However, the recombination centers are present in Rh-related defect configurations attributable to Rh localized orbitals in the forbidden region and moreover, there is a large shift in the CBm or VBM. This will lead to degradation in photocatalytic efficiency. The mobility of charge carriers is maximum in Mn$_\textrm{Ti}$S$_\textrm{O}$ codoped SrTiO$_3$, and in rest of the cases, it is not affected much. Our results reveal that Mn$_\textrm{Ti}$S$_\textrm{O}$ codoped, S$_\textrm{O}$ monodoped anatase TiO$_2$, Mn$_\textrm{Ti}$S$_\textrm{O}$ and Mn$_\textrm{Sr}$N$_\textrm{O}$ codoped SrTiO$_3$ are the most favorable candidates for enhancing photocatalytic overall water splitting owing to the passivation of trap states and congenial band edge positions with desirable visible light absorption. 

\section*{Acknowledgements}
MK acknowledges CSIR, India, for the senior research fellowship [grant no. 09/086(1292)/2017-EMR-I]. PB acknowledges UGC, India, for the senior research fellowship [grant no. 20/12/2015(ii)EU-V]. SS acknowledges CSIR, India, for the senior research fellowship [grant no. 09/086(1231)/2015-EMR-I]. SB acknowledges the financial support from SERB under core research grant (grant no. CRG/2019/000647). We acknowledge the High Performance Computing (HPC) facility at IIT Delhi for computational resources.

%%%REFERENCES%%%
\bibliography{sci} %You need to replace "rsc" on this line with the name of your .bib file
\end{document}

\title{\textbf{\Large{Theoretical insights of codoping to modulate electronic structure of TiO$_2$ and SrTiO$_3$ for enhanced photocatalytic efficiency}}}
	\author{Manish Kumar$^*$, Pooja Basera, Shikha Saini, Saswata Bhattacharya$^*$\\Department of Physics, Indian Institute of Technology Delhi, New Delhi 110016 India \\
		$^*$Email: manish.kumar@physics.iitd.ac.in [MK], saswata@physics.iitd.ac.in [SB]}
	%\date{\today}
	\pacs{}
	%\keywords{CH$_3$NH$_3$PbI$_3$, CH$_3$NH$_3$SnI$_3$, defects, DFT, vacancy, carrier concentration, free energy}
	\maketitle
	%%%%%%%%%%%%%%%%%
\begin{center}
	{\Large \bf Supplemental Material}\\ 
\end{center}
\begin{enumerate}[\bf I.]
	\item Electronic density of states (DOS) of doped anatase TiO$_2$ and SrTiO$_3$
	\item Optical properties of monodoped anatase TiO$_2$ and SrTiO$_3$
	\item Band edge alignment of monodoped anatase TiO$_2$ and SrTiO$_3$
\end{enumerate}
%\vspace*{12pt}
%\clearpage
\newpage
%
% Uncomment for keywords
%\vspace{2pc}
%\noindent{\it Keywords}: metal oxide, TiO$_2$, SrTiO$_3$, defects, formation energy, optical absorption, water splitting
%
% Uncomment for Submitted to journal title message
%\submitto{J. Phys.: Mater.}
%
% Uncomment if a separate title page is required
%\maketitle
% 
% For two-column output uncomment the next line and choose [10pt] rather than [12pt] in the \documentclass declaration
%\ioptwocol
%
\section{E\MakeLowercase{lectronic density of states} (DOS) \MakeLowercase{of doped anatase} T\MakeLowercase{i}O$_2$ \MakeLowercase{and} S\MakeLowercase{r}T\MakeLowercase{i}O$_3$}
\begin{figure}[H]
	\centering
	\includegraphics[width=0.8\textwidth]{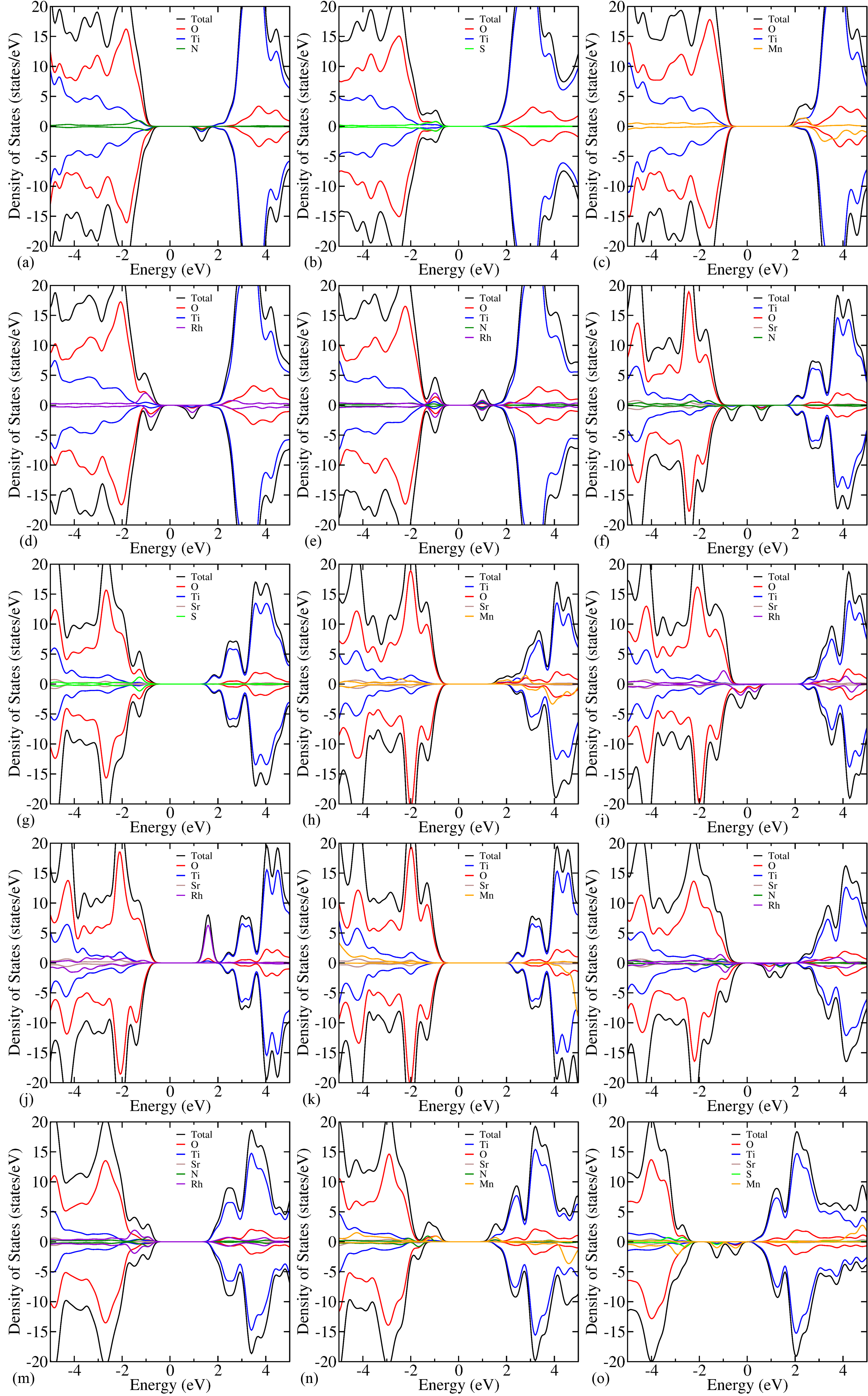}
	\caption{Atom projected density of states of (a) N$_\textrm{O}$, (b) S$_\textrm{O}$, (c) Mn$_\textrm{Ti}$, (d) Rh$_\textrm{Ti}$ (e) Rh$_\textrm{Ti}$N$_\textrm{O}$ codoped TiO$_2$, and (f) N$_\textrm{O}$, (g) S$_\textrm{O}$, (h) Mn$_\textrm{Ti}$, (i) Rh$_\textrm{Ti}$, (j) Rh$_\textrm{Sr}$, (k) Mn$_\textrm{Sr}$, (l) Rh$_\textrm{Ti}$N$_\textrm{O}$, (m) Rh$_\textrm{Sr}$N$_\textrm{O}$, (n) Mn$_\textrm{Sr}$N$_\textrm{O}$ %~\cite{jpc_sto}
	and (o) Mn$_\textrm{Sr}$S$_\textrm{O}$ codoped SrTiO$_3$.}
	\label{fig_DOS}
\end{figure}
\section{O\MakeLowercase{ptical properties of monodoped anatase} T\MakeLowercase{i}O$_2$ \MakeLowercase{and} S\MakeLowercase{r}T\MakeLowercase{i}O$_3$}
\begin{figure}[H]
	\centering
	\includegraphics[width=0.8\columnwidth]{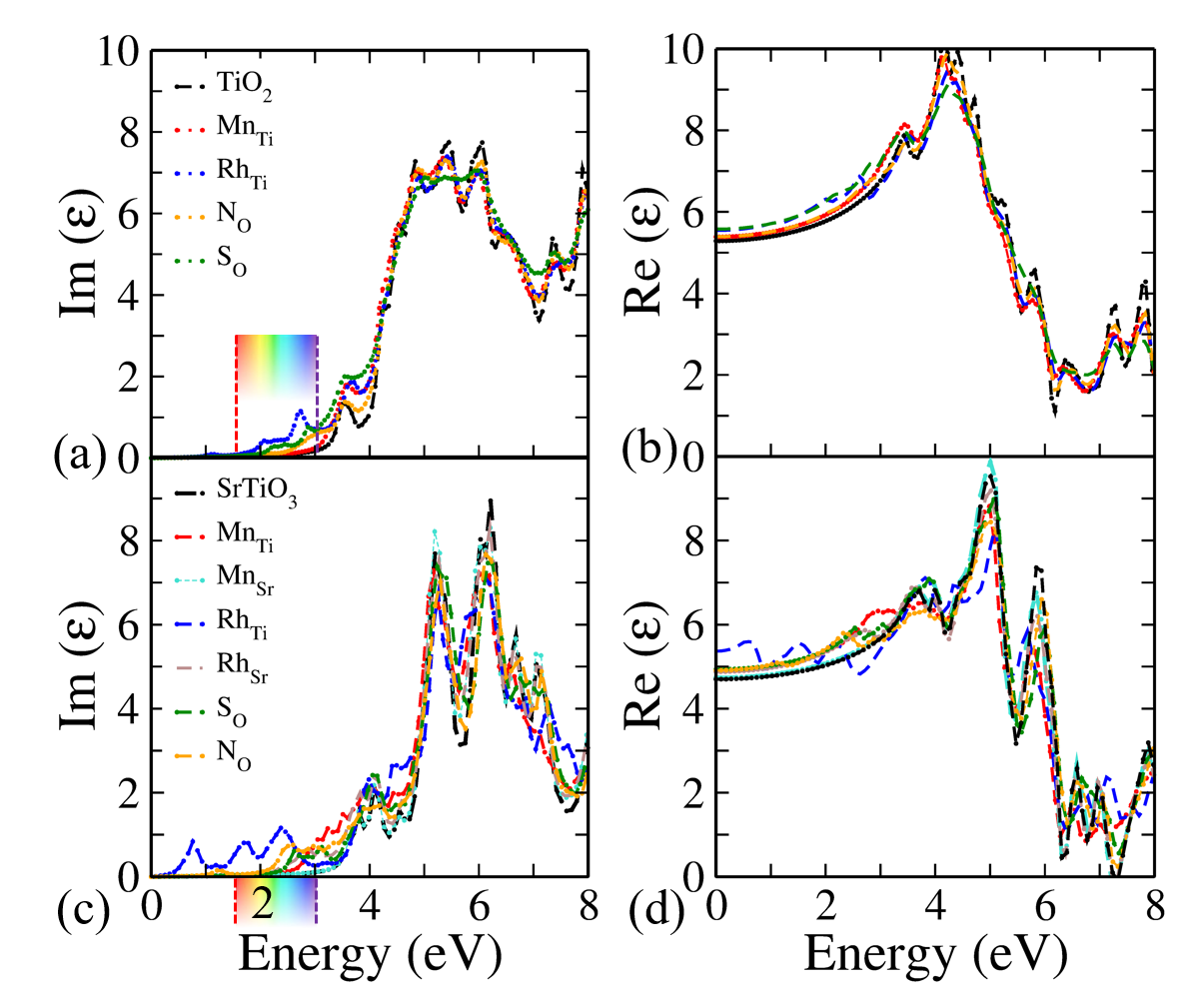}
	\caption{Spatially average (a) imaginary (Im $\varepsilon$) and (b) real (Re $\varepsilon$) part of the dielectric function for (un)doped TiO$_2$, (c) imaginary (Im $\varepsilon$) and (d) real (Re $\varepsilon$) part for (un)doped SrTiO$_3$.}
	\label{fig_opt}
\end{figure}
\section{B\MakeLowercase{and Edge Alignment of monodoped anatase} T\MakeLowercase{i}O$_2$ \MakeLowercase{and} S\MakeLowercase{r}T\MakeLowercase{i}O$_3$}
\begin{figure}[H]
	\centering
	\includegraphics[width=0.8\textwidth]{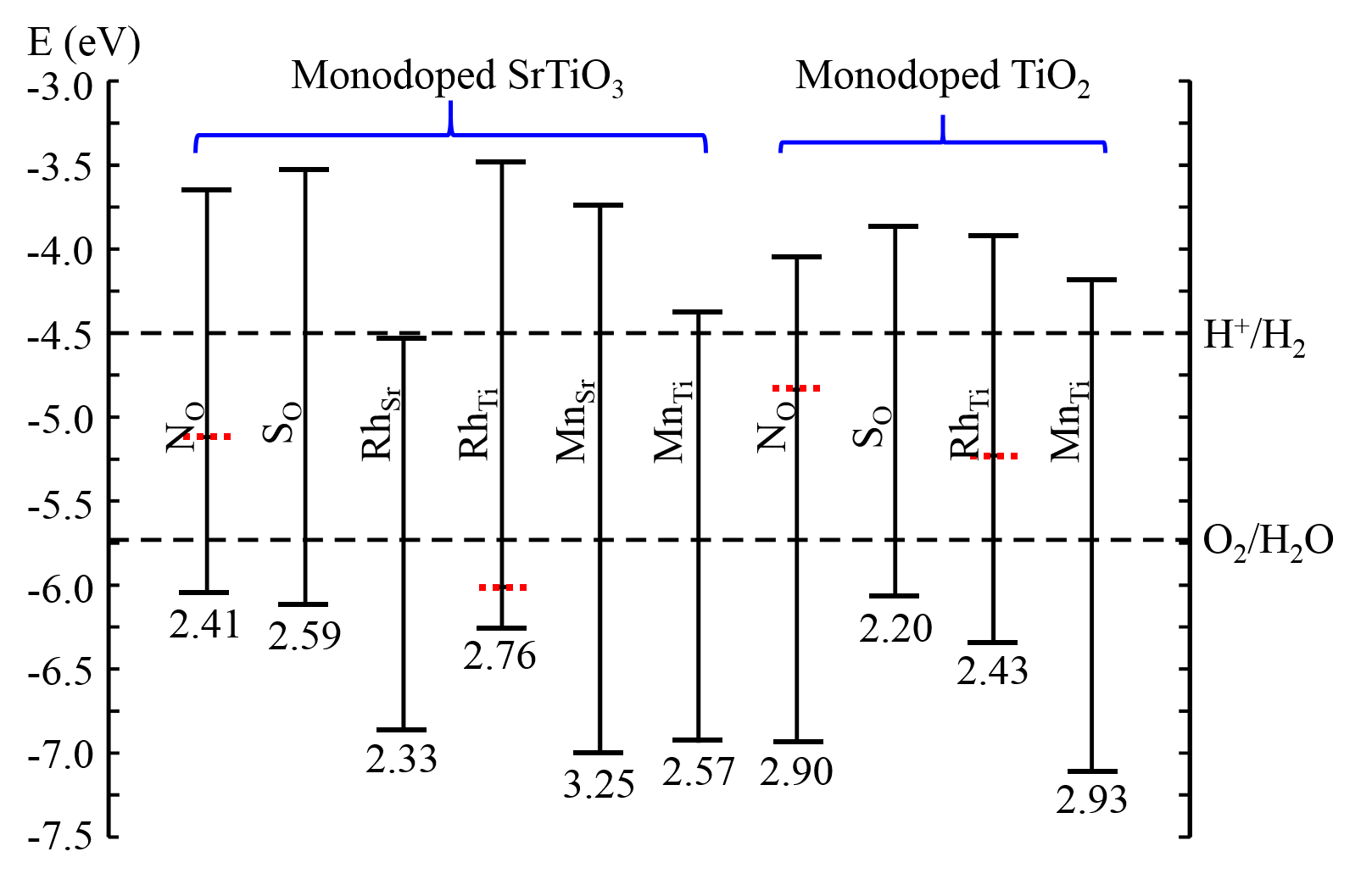}
	\caption{Band edge alignment of monodoped SrTiO$_3$ and TiO$_2$ w.r.t. water reduction and oxidation potential levels (H$^+$/H$_2$, O$_2$/H$_2$O). The dashed red line in forbidden region is representing the lowest unoccupied states.}
	\label{fig_band_edge_align}
\end{figure}
%\section*{References}
%\begin{thebibliography}{}
%	\bibitem{jpc_sto}
%	Kumar M, Basera P, Saini S and Bhattacharya S 2019 to be published
%\end{thebibliography}{}